\documentclass[superscriptaddress,twocolumn]{revtex4}

\usepackage{graphicx}
\usepackage[normalem]{ulem}
\usepackage{natbib}
\usepackage{amsmath,array,amssymb}
\usepackage{dcolumn}
\usepackage{bm}
\usepackage{hyperref}
\usepackage{color}
\usepackage{amsmath, amsthm, amssymb, amsfonts}

\graphicspath{{img/}}

\begin{document}
	
	\title{Storage and retrieval of heralded telecommunication-wavelength photons using a solid-state waveguide quantum memory}

	\author{Mohsen Falamarzi Askarani}
	\affiliation{Institute for Quantum Science and Technology, and Department of Physics \& Astronomy, University of Calgary, 2500 University Drive NW, Calgary, Alberta, T2N 1N4, Canada}
	
	\author{Marcel.li Grimau Pugibert}	
	\affiliation{Institute for Quantum Science and Technology, and Department of Physics \& Astronomy, University of Calgary, 2500 University Drive NW, Calgary, Alberta, T2N 1N4, Canada}
	\affiliation{current address: University of Basel, Klingelbergstrasse 82, CH-4056 Basel, Switzerland}
	
	\author{Thomas Lutz}
	\affiliation{Institute for Quantum Science and Technology, and Department of Physics \& Astronomy, University of Calgary, 2500 University Drive NW, Calgary, Alberta, T2N 1N4, Canada}
	\affiliation{current address: ETH Z{\"u}rich, Otto-Stern-Weg 1, 8093 Zürich, Switzerland}
	\author{Varun B. Verma}
	\affiliation{National Institute of Standards and Technology,
		325 Broadway, Boulder, Colorado 80305, USA}
	\author{Matthew D. Shaw}
	\affiliation{Jet Propulsion Laboratory, California Institute of Technology, 4800 Oak Grove Drive, Pasadena, California 91109, USA}
	\author{Sae Woo Nam}
	\affiliation{National Institute of Standards and Technology,
		325 Broadway, Boulder, Colorado 80305, USA}
	\author{Neil Sinclair}
	\affiliation{Institute for Quantum Science and Technology, and Department of Physics \& Astronomy, University of Calgary, 2500 University Drive NW, Calgary, Alberta, T2N 1N4, Canada}
	\affiliation{current address: California Institute of Technology, 1200 East California Boulevard, Pasadena, California 91125, USA}
	\author{Daniel Oblak}
	\affiliation{Institute for Quantum Science and Technology, and Department of Physics \& Astronomy, University of Calgary, 2500 University Drive NW, Calgary, Alberta, T2N 1N4, Canada}
	\author{Wolfgang Tittel}
	\affiliation{Institute for Quantum Science and Technology, and Department of Physics \& Astronomy, University of Calgary, 2500 University Drive NW, Calgary, Alberta, T2N 1N4, Canada}
	\affiliation{current address: QuTech, Delft University of Technology, 2600 GA Delft, The Netherlands}
	\date{\today}
	
	\begin{abstract}
	Large-scale quantum networks will employ telecommunication-wavelength photons to exchange quantum information between remote measurement, storage, and processing nodes via fibre-optic channels. Quantum memories compatible with telecommunication-wavelength photons are a key element towards building such a quantum network. Here, we demonstrate the storage and retrieval of heralded 1532 nm-wavelength photons using a solid-state waveguide quantum memory. The heralded photons are derived from a photon-pair source that is based on parametric down-conversion, and our quantum memory is based on a 6 GHz-bandwidth atomic frequency comb prepared using an inhomogeneously broadened absorption line of a cryogenically-cooled erbium-doped lithium niobate waveguide. Using persistent spectral hole burning under varying magnetic fields, we determine that the memory is enabled by population transfer into niobium and lithium nuclear spin levels. Despite limited storage time and efficiency, our demonstration represents an important step towards quantum networks that operate in the telecommunication band and the development of on-chip quantum technology using industry-standard crystals.
	\end{abstract}
	
	
	\maketitle
	Many efforts towards future quantum networks \cite{kimble2008quantum} have focused on employing photons at wavelengths in the C band (1530-1565 nm) due to the possibility of low-loss transmission using existing fiber-optic telecommunication infrastructure. Local nodes in a quantum network are envisioned to process information using (optical) chips that are comprised of (elementary) quantum computers \cite{kimble2008quantum}, while the synchronization of information is enabled by quantum memories that store and retrieve quantum information \cite{lvovsky2009optical}. The latter underpins the operation of quantum repeaters, which promise to transmit quantum information through lossy channels or over intercontinental distances \cite{sangouard2011quantum}. 
	
	Efforts to realize long-distance quantum communication have been bolstered by the significant progress of researchers to develop of quantum-optical technology \cite{2009photonicQtechnology}. However, much of this work is not compatible with C-band photons without the use of optical frequency conversion \cite{tanzilli2005photonic}. To avoid the conversion step, efforts have focused towards quantum technology that operates in the C-band, which has resulted in the development of efficient photon detectors \cite{natarajan2012SNSPD} and single-photon sources \cite{Quantumdots}. Nonetheless, a C-band quantum memory has turned out to be particularly challenging component to realize and in particular so if it is to be integrated into a solid-state platform such that it is easy to integrate with other photonics components. 
	
	Significant progress towards quantum memories has been made in the last decade, with most focus on the development of optical, microwave or radio-frequency to matter interfaces \cite{bussieres2013prospective}. One promising approach to implement quantum memory is based on cryogenically-cooled rare-earth-ion-doped crystals as they often feature long optical and spin coherence times, as well as suitable energy-level structures \cite{Charles2011REI}. Achievements with rare-earth-ion-doped crystals include broadband storage of entangled photons \cite{saglamyurek2011broadband,clausen2011quantumcrystal}, teleportation into quantum memories \cite{bussieres2014teleportationintomemory}, storage in nano-fabricated structures\cite{zhong2017on-chipstorage}, and storage assisted by impedance-matched cavities \cite{sabooni2013cavity,afzelius2014cavity}. Furthermore, rare-earth-ion-based quantum memory and signal processing has been demonstrated using industry-standard waveguides, such as titanium-indiffused lithium niobate (Ti$^{4+}$:LiNbO$_{3}$) \cite{saglamyurek2011broadband,saglamyurek2014manipulation,sinclair2017properties}, which are promising for efficient on-chip information processing.
	
	Er-doped crystals are unique in that they offer a ground-to-excited level transition in the C band with a coherence time that is the longest of any optical transition in a solid \cite{bottger2003erbium}. This property, found in Er-doped yttrium-orthosilicate (Er:Y$_2$SiO$_5$), prompted the realization of a photon-echo memory protocol for coherent light at the single-photon level \cite{2010YSiOmemorynoise}, but not for non-classical (single photon) light.
	A disadvantage of most Er-doped materials, including Er:Y$_2$SiO$_5$, is the relatively long lifetime of the excited level (up to $\approx$11 ms) as compared to that of sub-levels of the ground level ($\approx$130 ms), which makes optical pumping to the latter inefficient. This poses a significant obstacle to achieve a high-efficiency in photon-echo based quantum memory protocols, which rely on the optical pumping of sub-ensembles to shelving levels such as the electronic sub-levels of the ground state. This obstacle is partially overcome in Er-doped fibers, which due to their amorphous (rather than crystalline) structure feature reduced spin-spin interaction strengths compared to crystals and, as a result, extended ground-level lifetimes \cite{hole-erbium}. This feature allowed a recent demonstration of quantum memory for non-classical and entangled states of light at telecommunication wavelengths \citep{saglamyurek2015quantum}. Despite the appeal of Er-doped fibres for all-fiber implementations, decoherence arising from their amorphous structure \cite{hole-erbium,Coherence-fiber} currently restricts the resulting memories' storage times to less than one hundred nanoseconds, though improvements at ultra-low temperatures may be possible \cite{saglamyurek2016multiplexed}. Thus, the demonstration of a telecommunication-wavelength quantum memory for non-classical light using a solid-state crystal, in particular a crystalline waveguide that can be integrated with telecommunication-industry devices (such as modulators that are realized with Ti$^{4+}$:LiNbO$_{3}$ waveguides), is an important and yet-to-be achieved goal.
	
	Here we demonstrate a quantum memory for heralded single photons at 1532.05~nm wavelength using an erbium- and titanium-indiffused lithium-niobate waveguide (Er$^{3+}$:Ti$^{4+}$:LiNbO$_{3}$) (see Methods for fabrication details) cooled to 0.6~K. By spectrally-selective optical pumping to shelving levels, which we identify as arising from the (superhyperfine) coupling of $^{7}$Li and $^{93}$Nb nuclear spins in LiNbO$_{3}$ to an external magnetic field, we prepare a 6~GHz-bandwidth Atomic Frequency Comb (AFC) quantum memory. Heralded single-photons generated from a spontaneous parametric down conversion (SPDC) based photon pair-source are stored and retrieved from the AFC. The quantum nature of the recalled heralded photons is verified by their cross-correlation value with the heralding photon.
	
	\subsection*{Experiment}
	Our quantum memory is based on the AFC protocol, which is well suited to the spectroscopic properties of rare-earth ion doped materials and has underpinned much progress towards efficient and broadband quantum memories \cite{liu2006sREIs-book}. For example, storage times as long as milliseconds \cite{AFC2015ms}, efficiencies as high as 56\% \cite{sabooni2013cavity}, storage bandwidths of several GHz \cite{saglamyurek2011broadband} and fidelities up to 99.9\% \cite{Guo2012realization} have been achieved. An AFC is comprised of a series of spectral absorption lines that are equidistantly detuned by $\Delta$ (i.e. a 'comb' of absorption features) \cite{unitefficiency1}.
	An AFC can be prepared on the inhomogeneously broadened absorption line of rare-earth ions by spectrally-selective optical pumping into long-lived sub-levels (often hyperfine ground-levels). After preparation, an incident photon is absorbed by the comb to generate a collective atomic excitation that is described by
	\begin{equation}
	\left|\Psi\right>_{A}=\frac{1}{\sqrt{N}}\sum_{j=1}^N c_{j}e^{i2\pi\delta_{j}t} e^{-ikz_{j}}\left|g_{1}, ...e_{j},...g_{N}\right>,
	\label{equ:AFC}
	\end{equation}	
	where $N$ is the number of ions and $\delta_{j}$ is the detuning of the $j$th atom's transition frequency with respect to the frequency of the incoming photon. The longitudinal position of the $j$th ion is denoted by $z_j$, and the coefficient $c_j$ is related to its excitation probability. After absorption, each term in Eq.~\eqref{equ:AFC} accumulates a different phase according to its detuning $\delta_j = m_j\Delta$, where $m_j$ is an integer. Consequently, all terms coherently rephase at a time $\tau = 1/\Delta$, yielding collective re-emission of the photon in its original quantum state. Under certain conditions the retrieval process can in principle reach unit efficiency \cite{unitefficiency1,unitefficiency2,unitefficiency3}. The experimental performance of the AFC quantum memory hinges on the spectroscopic properties of the rare-earth-ion doped crystal used. For example, the level structure and population dynamics influence the storage efficiency, operating wavelength as well as bandwidth \cite{lvovsky2009optical,Tittel2010reviewQM,bussieres2013prospective}. 
	
	\begin{figure*}[t]
		\centering
		\includegraphics[width=2\columnwidth]{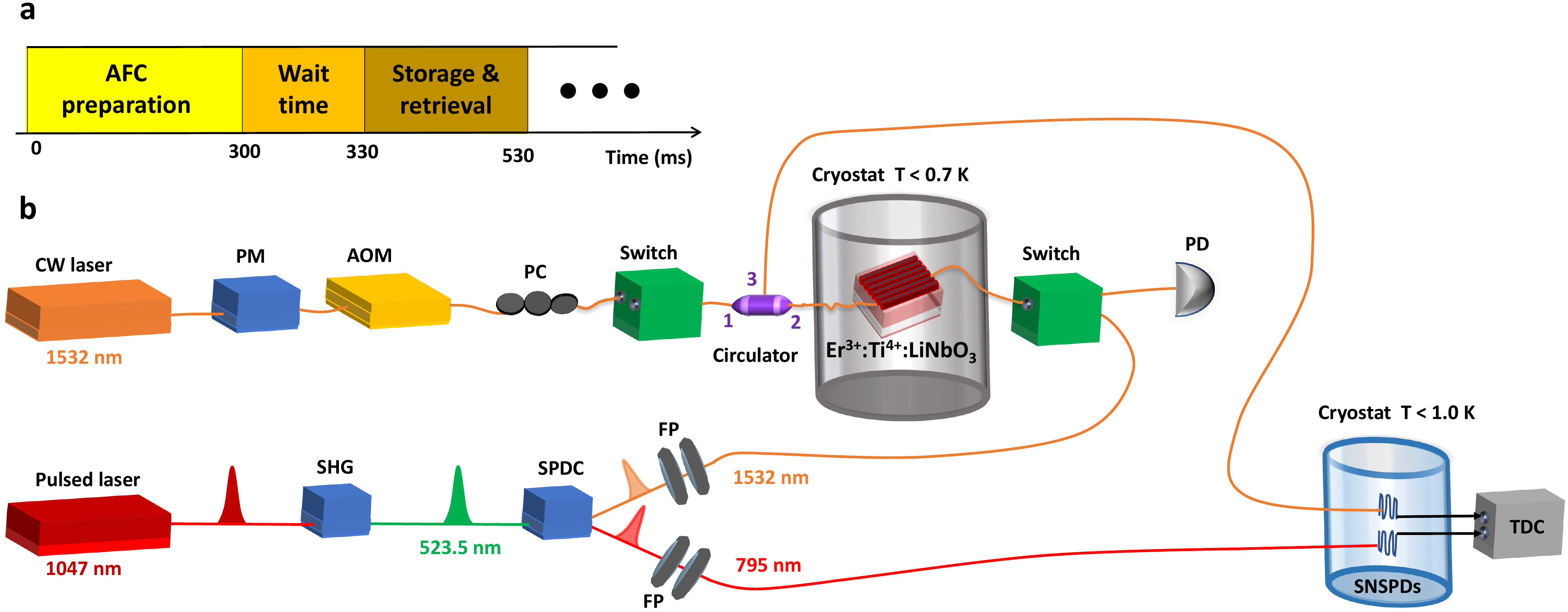}
		\caption{a) Experimental timing sequence for AFC preparation as described in the Methods. "$\cdot \cdot \cdot$" denotes repetition of the sequence. b) Experimental set-up as described in the main text and Methods consists of two main parts. Top half, which is for optical pumping for spectral hole burning measurements and AFC generation consists of a continuous wave laser (CW laser), phase modulator (PM), accousto-optic modulator (AOM), polarization controller (PC), optical switches, a circulator, and a photo-detector (PD) which is used to record the optical depth spectra shown in Figs.~\ref{fig:Super-hyper fine splitting}a and \ref{fig:AFC}. Bottom half, which is for heralded single photon generation and detection consists of a pulsed pump laser, second harmonic generation (SHG), spontaneous parametric down conversion (SPDC), Fabry-perot cavity (FP), superconducting nanowire single-photon detectors (SNSPD), time to digital converter (TDC).}
		\label{fig:setup}
	\end{figure*}
	
	There have been only a few spectroscopic studies of the low-temperature properties of the $^{4}I_{15/2}$ $\leftrightarrow$ $^{4}I_{13/2}$ telecom-wavelength transition in our material of choice, Er$^{3+}$-doped LiNbO$_{3}$.
	\cite{Ahdiye2009,Charlesppt,thiel2010LNbulk,Charles2011REI,thiel2012LNcrystalprocessing,lutz2017powder}. The Er ion is a so-called Kramers ion and the electronic spin degeneracy is lifted by the Zeeman effect in an applied magnetic field. The result is a split doublet in both the excited and ground level of the $^{4}I_{15/2}$ $\leftrightarrow$ $^{4}I_{13/2}$ transition \cite{liu2006sREIs-book}. A ground-level electronic Zeeman splitting of 1.6~MHz/G was measured for a 0.005\%-doped Er-doped lithium niobate bulk crystal (Er$^{3+}$:LiNbO$_{3}$, not containing titanium) in magnetic field parallel to the c-axis of the crystal \cite{Charlesppt}. In addition, rare-earth ions can couple to nuclear spins of crystal lattice ions (superhyperfine coupling)\cite{liu2006sREIs-book}. For Er$^{3+}$:LiNbO$_{3}$, the electron spin (1/2) of erbium is known to couple to nearby lithium ($I=3/2$) and niobium ($I=9/2$) spins, and possibly other impurities \cite{Charlesppt}. A study of Er$^{3+}$:LiNbO$_{3}$ at temperatures as low as 1.6~K using light polarized orthogonal to the crystal's c-axis (i.e. E$\perp$c) revealed a 180~GHz-wide inhomogeneously-broadened line, a 2~ms population lifetime of the $^{4}I_{13/2}$ level, and an optical coherence lifetime of up to 117~$\mu$s at a 50~kG field oriented parallel to the c-axis (B$\parallel$c) \cite{thiel2010LNbulk,Charles2011REI}. The same transition of Er$^{3+}$ was studied in a Ti$^{4+}$:LiNbO$_{3}$ waveguide at 3~K with E$\perp$c-polarized light to reveal a 250~GHz-wide inhomogneously broadened line, a population lifetime matching that of the bulk material, and a coherence lifetime of 18.2~$\mu$s in a 4~kG field oriented B$\parallel$c \cite{Ahdiye2009}. Regardless, of the slightly degraded performance\footnote{We expect that at a lower temperature, optimized magnetic field, and with consideration of instantaneous spectral diffusion \cite{liu2006sREIs-book}, the coherence lifetime in the waveguide will approach that of the bulk, similarly to what was observed for Tm$^{3+}$:Ti$^{4+}$:LiNbO$_{3}$ \cite{sinclair2017properties}.} of the waveguide compared to  bulk Er$^{3+}$:LiNbO$_{3}$ under these conditions, the coherence properties suffice for up to $\mu$s-long storage times.
	
\begin{figure}
		\centering
		\includegraphics[width=1\columnwidth]{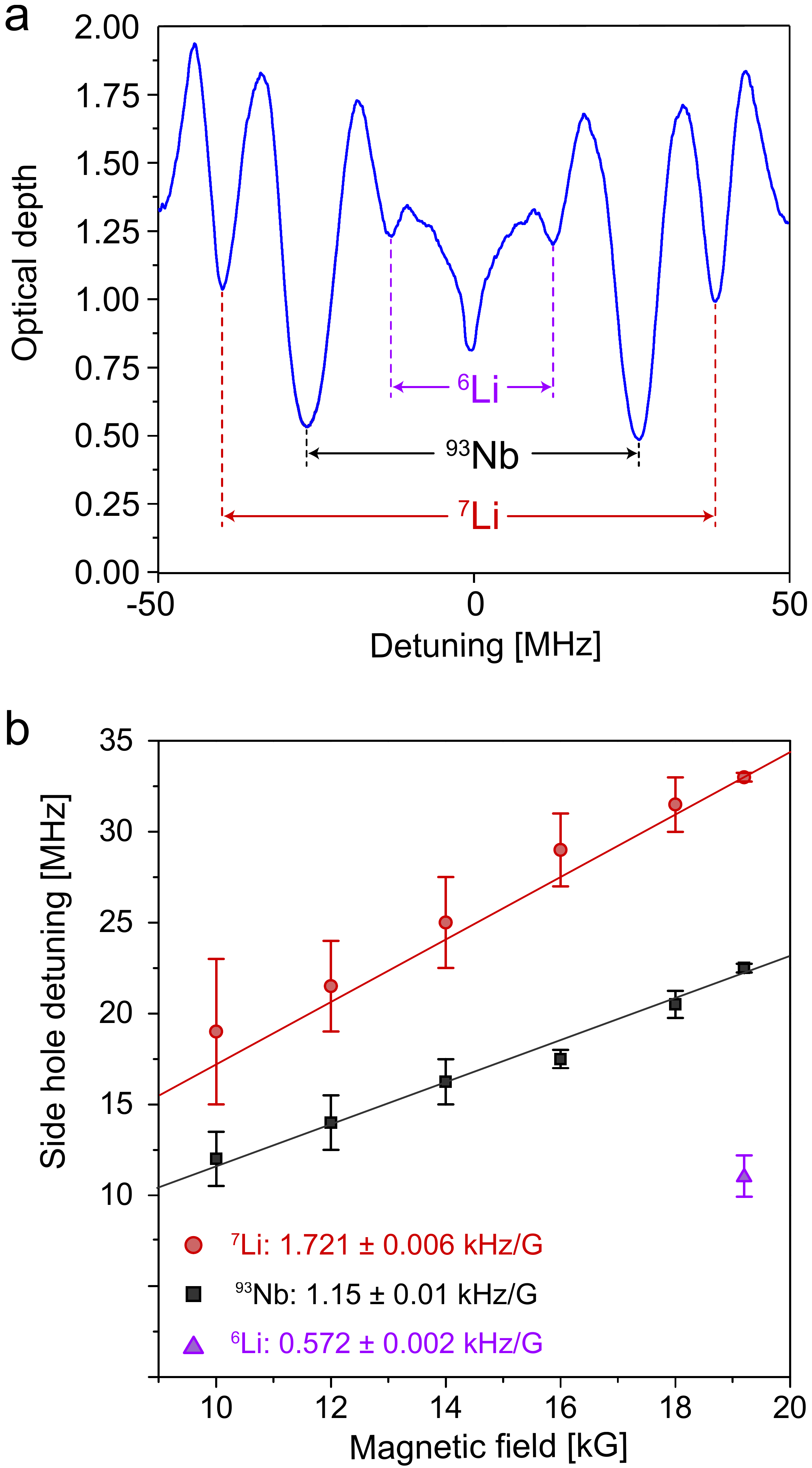}
		\caption{a) Optical depth over a 100~MHZ spectral range centered at 1532.05~nm in a 19~kG magnetic field showing evidence of spectral hole burning. Side holes due to the interaction of the Er$^{3+}$ electronic spin with $^{93}$Nb, $^{6}$Li, and $^{7}$Li nuclear spins of LiNbO$_{3}$ are labeled accordingly. The apparent side-holes at $\pm$50~MHz are due to imperfect frequency sweep. b) Detuning of $^{7}$Li and $^{93}$Nb side-holes as function of magnetic field. Side-holes for $^{6}$Li are resolved only at the highest magnetic field. Uncertainty bars are calculated from the between the negative and positive detuned side-holes.}
		\label{fig:Super-hyper fine splitting}
	\end{figure}

	For our Er$^{3+}$:Ti$^{4+}$:LiNbO$_{3}$ waveguide to be suitable for the AFC protocol we additionally need to verify the existence shelving levels with a lifetime significantly exceeding that of the excited state. To this end we perform spectral hole burning to determine the structure and population dynamics of the hyperfine levels (i.e. electronic Zeeman and superhyperfine levels) \cite{hole-erbium} using the setup sketched in Fig.~\ref{fig:setup} with the crystal at 0.6~K. Fig.~\ref{fig:Super-hyper fine splitting}a depicts the resulting absorption spectrum (at a magnetic field of 19~kG), which displays a rich side- and anti-hole structure, which persists up to a time delay of 15~minutes. (Here and henceforth, the light polarization and the magnetic field orientation are set perpendicular and parallel to the c-axis of the crystal, respectively, i.e. E$\perp$c and B$\parallel$c.) As described in the Methods, the side-hole detunings are determined by the excited-level hyperfine structure, whereas the detunings of the anti-holes are given by the energy-level differences between the excited- and ground-level hyperfine structure \cite{Thierryside-antiholes}.

	To determine the origin of the substructure, hole burning is repeated at different magnetic fields, and the detuning of the central frequencies of each side-hole is measured after a time delay of 30~ms.
	We plot the field-dependent detuning in Fig.\ref{fig:Super-hyper fine splitting}b for three pairs of distinguishable side-holes. A fit to their field dependence gives values of 1.15$\pm$0.01, 1.721$\pm$0.006, and 0.572$\pm$0.002 kHz/G, respectively. This agrees well with the superhyperfine coupling of Er$^{3+}$ to $^{93}$Nb, $^{7}$Li, and $^{6}$Li nuclear spins in the host crystal as previously measured in Er$^{3+}$:LiNbO$_{3}$ \cite{1991nuclearspinsplitting,thiel2010LNbulk}.
	At fields below 10~kG the superhyperfine splitting is not resolvable, but we still observe long-lived spectral holes. These are likely due to optical pumping to long lived electronic Zeeman levels of Er$^{3+}$. However, no side or anti-hole structure is discernible and thus no Zeeman level splitting can be determined.
	
	To gain a better understanding of the population dynamics we perform time-resolved spectral hole burning using magnetic fields of 0.35, 0.6, 0.8 and 19~kG. Based on earlier measurements of the spectral diffusion in Er$^{3+}$:LiNbO$_{3}$ \cite{thiel2010LNbulk} we expect that the few MHz linewidth will dominate over any spectral diffusion in our material.
	For the largest field, we monitor the depth of the central hole after a variable time delay between $\sim$30~s and $\sim$20~min and estimate the $1/e$ hole lifetime to be 10 min. This agrees well with that observed using Er$^{3+}$:LiNbO$_{3}$ under similar conditions \cite{Charlesppt}, where the population trapping mechanism was attributed to $^7$Li nuclear spins i.e. a superhyperfine level. 
	For the three lower fields, we vary the time delay from 20~ms to 4.5~s and record the depth of the central hole feature. The hole-decay features two distinct exponential decays. The $1/e$ hole lifetime for the faster decay is field-independent at $\sim$60 ms, while for the slower second decay it is of 1.0, 1.36, and 2.44~s for the three field values, respectively. The relative weights of all decays do not vary with field (see upcoming manuscript \cite{upcoming}). 
	We conjecture that the fast decay arises from relaxation of Er electronic Zeeman levels or superhyperfine levels. The former is supported by the short lifetimes of Zeeman levels that have been observed in many Er-doped crystals at these magnetic fields, including Er$^{3+}$:LiNbO$_{3}$.
	As for the latter, even though no clear superhyperfine structure is visible, the levels may be buried within the laser linewidth-broadened hole and thus residual optical pumping into superhyperfine levels could be possible.
	We attribute the long, magnetic-field dependent decay to relaxation of Zeeman levels of Er ions that belong to different magnetic classes, in the sense that their electronic Zeeman splitting is significantly different from neighboring Er ions. As a result of this electronic spin detuning, the spin flip-flop relaxation mechanism, which is the dominant one at these fields, is suppressed.
	Hence, relatively long Zeeman level lifetimes can be achieved similar to that in Er-doped fibers \cite{hole-erbium}. 
	Such spin-inhomogeneous broadening will be manifested by very broad anti-holes, which is corroborated by our inability to identify any anti-holes in the hole burning spectra. Further support for our interpretation comes from studies of Tm$^{3+}$-doped LiNbO$_{3}$ bulk crystal and waveguide \cite{sinclair2018,thiel2012LNcrystalprocessing}, in which significant anti-hole broadening caused by a distribution of Tm nuclear spin splittings was measured. This was attributed to the magnetic disorder of LiNbO$_{3}$ (e.g. due to Li$^{+}$-Nb$^{+5}$ anti-sites, as well as Nb$^{+5}$ and O vacancies). Further hole burning studies are necessary to shed light on these conjectures.
	
\begin{figure}
		\centering
		\includegraphics[width=0.85\columnwidth]{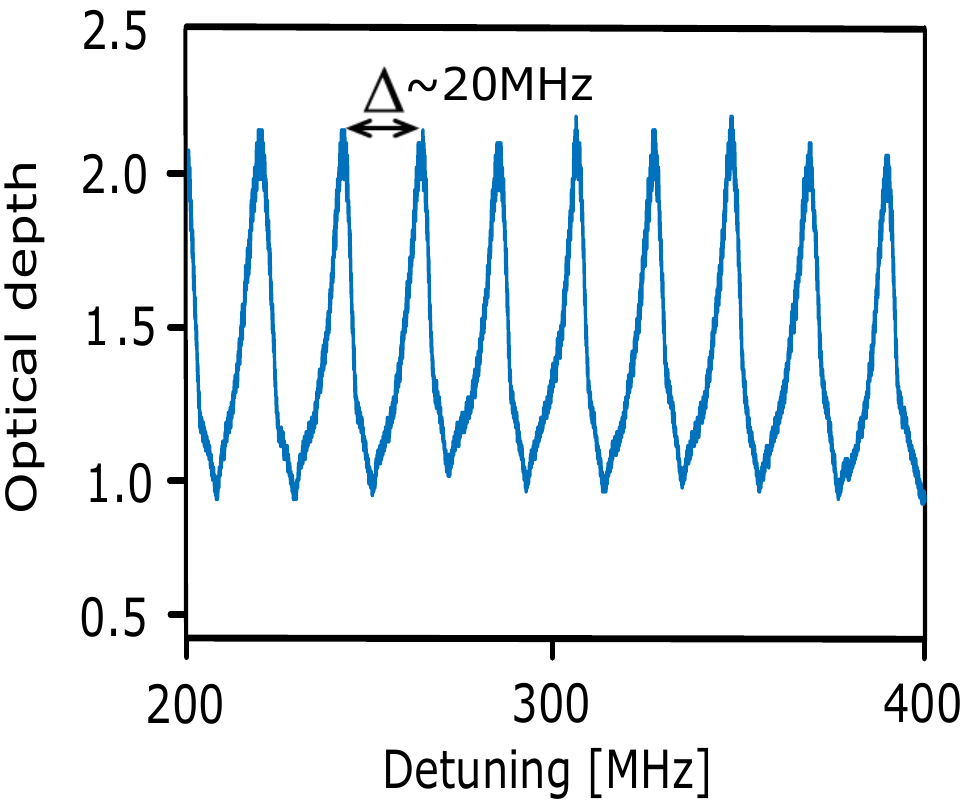}
		\caption{A 200 MHz-bandwidth section of our 6 GHz-bandwidth AFC.}
		\label{fig:AFC}
	\end{figure}

	Based on our hole burning measurements, it should be possible to realize the AFC by exploiting the long-lived population trapping mechanisms at either low or high fields.
	Nevertheless, we find that the AFCs generated at low fields are marred by a large absorption background when the bandwitdh is increased beyond about a GHz \cite{upcoming}. Hence, the high-field configuration with trapping in superhyperfine levels turns out to be advantageous. 
	We perform the optical pumping such that the comb periodicity somewhat coincides with the excited-level splitting caused by interactions with $^{7}$Li and $^{93}$Nb, which for the chosen magnetic field of 16.5 kG, are 28~MHz and 19~MHz, respectively. Based on this we chose an AFC storage time of $\tau=48~ns$. A 200 MHz-bandwidth section of our 6~GHz-bandwidth AFC structure is shown in Fig.~\ref{fig:AFC}. From the trace we expect an efficiency close to 1$\%$. However, limited laser intensity at large detunings leads to a non-uniform AFC and an averaged efficiency of around 0.3\%. The absorption background is due to imperfect hole burning. The finesse -- the ratio of tooth width $\gamma$ to tooth spacing $\Delta$ -- is two.

	\subsection*{Results and discussion}
\begin{figure}
		\centering
		\includegraphics[width=0.95\columnwidth]{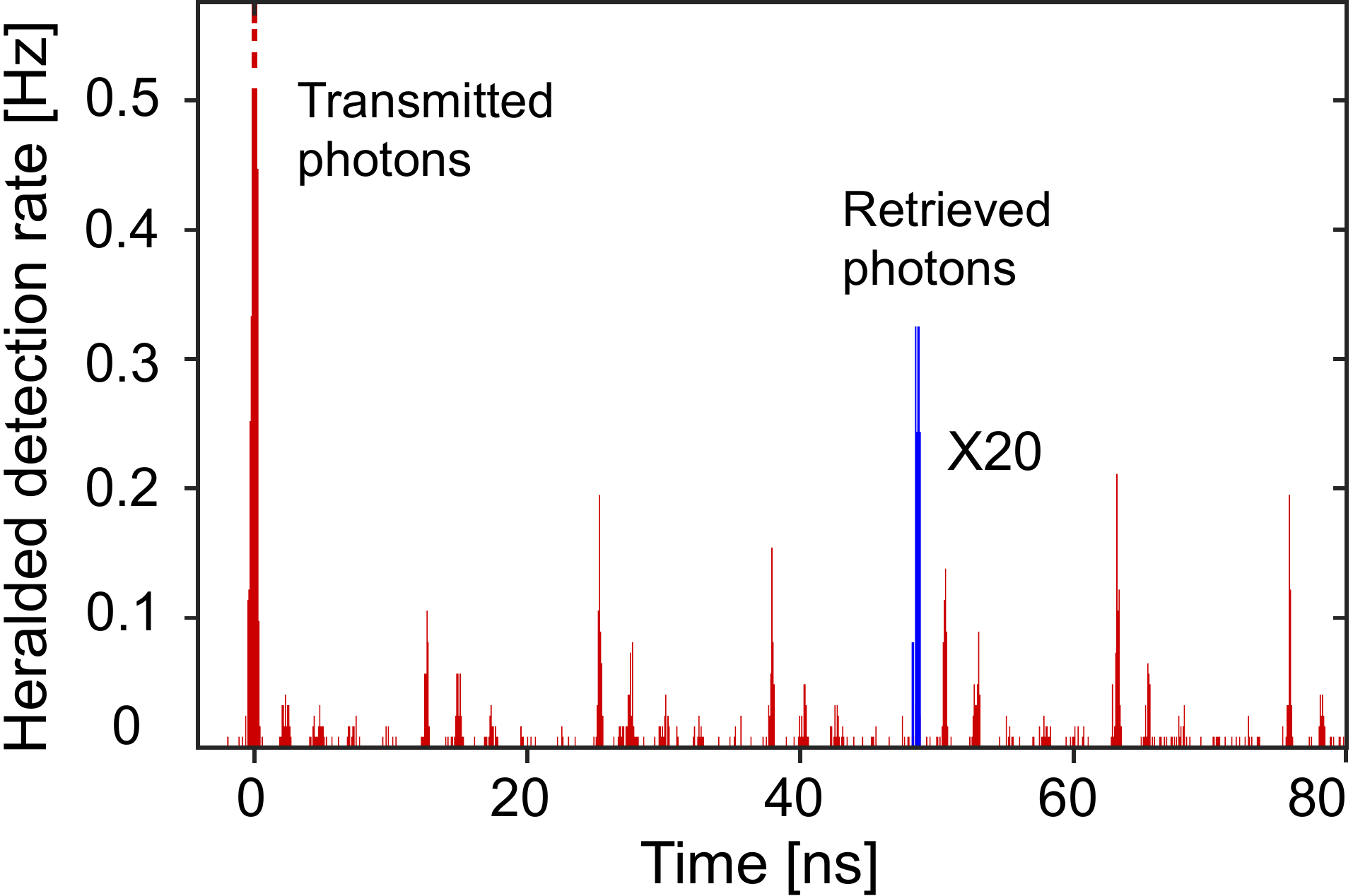}
		\caption{Quantum memory for a heralded telecommunication-wavelength photons. The histogram depicts the time-resolved signal-idler coincidence detection rate. The time axis is defined relative to the detection of the transmitted idler photons. The photons that are emitted by the AFC (colored in blue, magnified 20$\times$) are measured after a storage time of 48~ns. The sequence of coincidence peaks that repeat at a rate of 80 MHz is due to repetitive pumping of the SPDC source (explained in the main text).}
		\label{fig:echo}
	\end{figure}

	To demonstrate that our waveguide can be used as a quantum memory, we create and store (absorb) heralded telecommunication-wavelength photons of about 6~GHz bandwith, which are then re-emitted after a pre-programmed storage time of 48~ns. The experimental setup is outlined in Fig.~\ref{fig:setup} and the components described in detail in the Methods. Fig. \ref{fig:echo} depicts the histogram (red colour) of heralded detection events after the memory. The large peak $t=0$ corresponds to photons that are directly transmitted through our waveguide and thus not absorbed by the AFC structure. At $t=48$~ns the peak corresponding to stored and retrieved photons appears (in Fig.~\ref{fig:echo} it is magnified by a factor of 20 and coloured blue). The sequence of detection peaks separated by 12.5~ns are due to uncorrelated photons generated during different cycles of the 80~MHz repetition rate SPDC pump laser (often referred-to as accidental coincidences). Finally, the large number of smaller peaks in the histogram are due to spurious temporal modes of the pump laser -- a consequence of imperfect alignment of the laser cavity -- causing further uncorrelated photon-pair production events in the SPDC crystal.
	
	In order to assess the non-classical nature of the photons recalled from our quantum memory, we measure the second-order cross-correlation function g$^{(2)}$ extracted from the coincidence histogram (Fig. \ref{fig:echo}). This function is defined as $g^{(2)} = \frac{P_{SI}}{P_{S} \times P_{I}}$.  Here $P_{SI}$ corresponds to the probability of detecting a coincidence between the signal and idler photons, and $P_{S}$ ($P_{I}$) is the probability of detecting the signal (idler) photon. To prove that quantum correlations between the two photons are present, g$^{(2)}$  must be greater than 2 \cite{tapster1998photon}. First, bypassing the quantum memory, we  determine g$^{(2)}$ of our photon pair source (without storage) to be around 20. It is limited by multi-pair emissions caused by the high pump intensity \cite{tapster1998photon}. After one photon has been stored in our quantum memory, we measure a g$^{(2)}$ of 7.1$\pm$4, demonstrating that non-classical correlations between the members of photon pairs are still present after storage. 
	
	The memory efficiency is limited by several factors compared to the ideal efficiency of 54\% for forward recall \cite{unitefficiency1}, including the remaining absorption background, low finesse ($\Delta/\gamma$) of the AFC (where $\Delta$ is the tooth spacing and $\gamma$ the linewidth of each tooth) and, if $\gamma$ were reduced to increase the finesse and approach the maximum efficiency, insufficient optical depth. The limited optical depth and the lack of phase matching for backwards recall (in which case the efficiency can reach 100\%) can be remedied with an impedance-matched cavity \cite{unitefficiency2}. However, this solution is not beneficial without also reducing the remaining background loss, which we attribute, in our demonstration, to the complexity of the superhyperfine level structure and the possibility of laser-induced enhancement of spin relaxation \cite{thiel2012LNcrystalprocessing,sinclair2018}, an effect that has also been observed, and mitigated for the creation of AFCs in Tm$^{3+}$:Ti$^{4+}$:LiNbO$_{3}$ \cite{sinclair2017properties,sinclair2018}. In particular, the presence of (broad) anti-holes within the AFC troughs, which is possible for a host with considerable disorder such as LiNbO$_{3}$ \cite{thiel2012LNcrystalprocessing}, may not allow complete removal of the absorption background. The finesse of our AFC is currently limited to two by our optical pumping method \cite{unitefficiency1}. It is ultimately limited by the homogeneous linewidth of the used transition.
	
	We believe that further characterization of the atomic level structure and dynamics of the material, in particular through detailed hole burning measurements (e.g. with varying magnetic field strength and orientation), as well as optimization of the AFC preparation steps (e.g. using back-pumping or spin-mixing methods), the background absorption can be significantly reduced. To this end, the origin of the aforementioned long-lived holes at low magnetic fields should be studied further \cite{upcoming}. Furthermore, wavelength-dependent measurements may allow the identification of more favorable (super- or Zeeman-) hyperfine levels due to different magnetic sub-classes of ions. Furthermore, a lower sample temperature and concentration could allow for more efficient spectral hole burning in addition to suppressing decoherence mechanisms (e.g. spin flip-flops).
	
	Note that our current quantum memory implementation does not allow photons to be retrieved on demand, which requires in an additional step the reversible, coherent, mapping to a third long-lived level \citep{AFC_spinwave}. Nonetheless, applications such as quantum repeaters may utilize memories of fixed storage times provided that spectral or spatial multiplexing is utilized \cite{NeilPLRmultiplexing}.
	
	\subsection*{Conclusion}
	We have demonstrated the storage and retrieval of heralded 1532 nm telecommunication-wavelength photons using a cryogenically-cooled Er-doped lithium niobate waveguide. The non-classical nature of the storage process is demonstrated by a measurement of the cross-correlation between the heralding and the retrieved photons. We employed an AFC quantum memory that is based on optical pumping of atomic population into superhyperfine ground levels, and we have detailed the limitations as well as possible improvements of our memory, in particular its limited storage efficiency. Our work is a step towards developing on-chip quantum technology in the C band using standard materials that are employed by the telecommunication industry, and will benefit quantum networks.
	
	\begin{center}
		\textbf{METHODS}
	\end{center}
	\subsection*{Er-doped lithium niobate waveguide}
	
	To fabricate the Er$^{3+}$:Ti$^{4+}$:LiNbO$_{3}$ waveguide, z-cut congruent LiNbO$_{3}$ is Er-doped over a length of 10 mm by indiffusion of a vacuum-deposited (electron-beam evaporated) 8 nm-thick Er layer at 1130 $^\circ$C for 150 h in an Ar-atmosphere. This step is followed by a post-treatment in O (1 h) to get a full re-oxidization of the crystal. Er substitutes for Li when incorporated into the LiNbO$_{3}$ lattice. The indiffusion results in a 3.6~$\times$ 10$^{19}$~cm$^{-3}$ near-surface concentration and a Gaussian concentration profile that features a 1/e penetration depth of 8.2 $\mu$m \cite{baumann1996LiNbOfabrication}. Next, the waveguide is created by indiffusion of Ti. To do this, a 98 nm-thick titanium layer is deposited on the Er-doped surface of the LiNbO$_{3}$ substrate using electron beam evaporation. From this layer, 7 $\mu$m-wide Ti strips are defined by photo-lithography and chemical etching, and subsequently in-diffused at 1060 $^\circ$C for 8.55 h. This process leads to a single-mode waveguide with a 4.5~$\times$~3~$\mu$m full-width-at-half-maximum intensity distribution for transverse magnetic-polarization. Note that an analogous procedure, which is detailed in Ref. \cite{saglamyurek2011broadband}, was used to fabricate the similar Tm$^{3+}$:Ti$^{4+}$:LiNbO$_{3}$ waveguide.
	
	In the experiments the Er$^{3+}$:Ti$^{4+}$:LiNbO$_{3}$ waveguide is mounted in an adiabatic diamagnetic refrigerator (ADR) cryostat kept below 0.6~K and the in and out coupling of light is achieved by fiber butt-coupling. The overall in-out coupling transmission of our cryogenic setup is 20$\%$ (measured at a wavelength of 1532.05~nm).
	
	\subsection*{Spectral hole burning and AFC preparation}
	Spectral hole burning is performed by shining continuous wave  (CW) light at zero detuning $\Delta$ and subsequently recording the optical depth (OD) profile in a relevant spectral region centred on $\Delta=0$. The setup for this is depicted in the top half of Fig.~\ref{fig:setup}b. First a continuous-wave laser is switched to the crystal -- by using an acousto-optic modulator (AOM) -- in order to optically-pump a few MHz-wide frequency band of the inhomogeneous absorption line (at $\Delta=0$ corresponding to 1532.05~nm) until transparency, i.e. a spectral hole is created. The OD profile is recorded after a given delay by switching on the AOM and sweeping the frequency of the light impinging on the crystal over 100~MHz using serrodyne phase-modulation. The transmitted light is then detected on a photo-detector. A magnetic field oriented B$\parallel$c of up to 20~kG can be applied using a solenoid magnet. 
	
	The AFC is generated in a similar fashion except measures have to be taken to store and absorb single photons. First step is optical pumping for 300 ms as shown in Fig. \ref{fig:setup}a. This involves the 1532.05 nm-wavelength CW optical pumping light is frequency-swept over a 6 GHz bandwidth by serrodyne phase-modulation while the intensity modulated using an AOM. In the OD spectrum this procedure creates features of high (low pump power intensity) and low (high pump power intensity) absorption, as is shown in Fig.~\ref{fig:AFC} for a magnetic field of B$\parallel$c=16.5~kG . Following this is a 30~ms wait time 
	in order to avoid the impact of noise-inducing spontaneously emitted photons from the $^{4}I_{13/2}$ excited-level. Finally, during a 200~ms storage and retrieval duration heralded single photons are stored and retrieved in the waveguide memory. The optical switches and circulator are employed to suppress any stray otpical pump light during the storage and retrieval of the heralded single photons (see Fig. \ref{fig:setup}a). 
	
	\subsection*{Spectral hole structure}
	The CW light incident on the waveguides, results in the transfer of atomic population to the $^{4}I_{13/2}$ excited level, which subsequently decays and is redistributed among the hyperfine levels of the $^{4}I_{15/2}$ ground state. The resulting absorption profile allows the determination of the number of sub-levels. Furthermore, the detuning of the various features from the center of the (main) spectral hole can be used to extract the coupling strength of the involved spins and to identify population-trapping mechanisms. In addition, time-resolved measurements of the absorption profile can give insight into population dynamics. Regions of increased (decreased) transparency that are detuned from the burning frequency are referred to as side- (anti-) holes. 
	
	\subsection*{SPDC-based heralded single photon source}
	Photon pairs are generated using SPDC, a process in which a pump photon is converted into a pair of photons (whose total energy and momentum equal that of the pump photon) by using a second-order nonlinearity \cite{tanzilli2001SPDC}. As shown in Fig. \ref{fig:setup}, our production of photon pairs begins with a mode-locked laser that emits 6~ps-duration pulses at a rate of 80 MHz at a wavelength of 1047~nm. These pulses are frequency-doubled to 523.5~nm by means of second harmonic generation using a periodically-poled lithium niobate (PPLN) crystal and are then used to pump a second PPLN crystal that performs SPDC to yield frequency-correlated photon pairs of 795~nm (signal) and 1532~nm (idler) wavelength. After passing through an interference filter that removes the remaining 523~nm light, a dichroic mirror spatially separates the photons from each pair. Finally, the bandwidth of each photon is set by filtering, using a 12- (6-) GHz bandwidth fiber-Bragg grating (Fabry-Perot filter) for the 1532 (795)~nm photon. The 1532 nm idler photon is coupled to the waveguide and then stored. After retrieval, it is detected by a superconducting nanowire single-photon detector (SNSPD) \cite{fmarsiliSNSPD}. The heralding 795~nm signal photon is directly detected by the SNSPD. All detection events are recorded by a time-to-digital converter.
	
	\subsection*{Acknowledgments}
	The authors thank Wolfgang Sohler, Mathew George, and Raimund Ricken for providing the Er$^{3+}$:Ti$^{4+}$:LiNbO$_{3}$ waveguide, Francesco Marsili for assistance with development of the SNSPDs, Jakob H. Davidson for help with aligning the waveguide, and Gustavo Amaral and Erhan Saglamyurek for useful discussions. This work is funded through Alberta Innovates Technology Futures (AITF), the National Science and Engineering Research Council of Canada (NSERC), and the Defense Advanced Research Projects Agency (DARPA) Quiness program. W.T. acknowledges funding as a Senior Fellow of the Canadian Institute for Advanced Research (CIFAR), and V.B.V. and S.W.N. acknowledge partial funding for detector development from the Defense Advanced Research Projects Agency (DARPA) Information in a Photon (InPho) program. Part of the detector research was carried out at the Jet Propulsion Laboratory, California Institute of Technology, under a contract with the National Aeronautics and Space Administration (NASA).

\end{document}